\documentclass[12pt]{article}

\usepackage{latexsym} 
\usepackage{amssymb}  
\usepackage{epsfig}       

\newcommand{\beq}{\begin{equation}}
\newcommand{\eeq}{\end{equation}}
\newcommand{\ba}{\begin{array}}
\newcommand{\bea}{\begin{eqnarray}}
\newcommand{\ea}{\end{array}}
\newcommand{\eea}{\end{eqnarray}}

\newcommand\comment[1]{ \hbox{[{\it Comment suppressed here.}\/]} }
\newcommand\hide[1]{}

\newcommand{\bx}{{\vec x}}
\newcommand{\by}{{\vec y}}
\newcommand{\bs}{{\vec s}}

\newcommand{\bp}{{\vec p}}

\newcommand{\pa}{\parallel}
\newcommand{\pe}{\perp}

\newcommand{\skipover}[1]{}

\makeatletter 

\def\appendix{\par                              
    \setcounter{section}{0}                     
    \setcounter{subsection}{0}
    \renewcommand{\theequation}{\Alph{section}.\arabic{equation}}
    \renewcommand{\thesection}{Appendix \Alph{section}}
}

\def\applabel#1{\@bsphack
  \protected@write\@auxout{}%
         {\string\newlabel{#1}{{\Alph{section}}{\thepage}}}%
  \@esphack}

\def\section{
\setcounter{equation}{0}        
\@startsection {section}{1}{\z@}{-3.5ex plus -1ex minus
 -.2ex}{2.3ex plus .2ex}{\large\bf}}
\renewcommand{\theequation}{\arabic{section}.\arabic{equation}}

\def\subsection{\@startsection{subsection}{2}{\z@}{-3.25ex plus -1ex minus
 -.2ex}{1.5ex plus .2ex}{\normalsize\bf}}

\def\subsubsection{\@startsection{subsubsection}{3}{\z@}{-3.25ex plus
 -1ex minus -.2ex}{1.5ex plus .2ex}{\normalsize}}

\makeatother   

\newsavebox{\eqlabel}

\makeatletter  
\newlength{\numblen}
\newsavebox{\eqnumb}
\def\@eqnnum{\savebox{\eqnumb}{\rm (\theequation)}%
\settowidth{\numblen}{\usebox{\eqnumb}}%
\makebox[\numblen][l]{\usebox{\eqnumb}~~~\usebox{\eqlabel}}}
\makeatother   

\newenvironment{equationwithlabel}[1]{ %
  \savebox{\eqlabel}{#1}
  \begin{equation}\label{#1} }{\end{equation}} 
\newcommand{\beql}[1]{\begin{equationwithlabel}{#1}}
\newcommand{\eeql}{\end{equationwithlabel}}

\begin{document}

\title{\bf Range of validity of\\ transport equations\\[1.ex]}

\author{
J\"urgen Berges\thanks{email: j.berges@thphys.uni-heidelberg.de}$\,$ and
Szabolcs Bors\'anyi\thanks{email: s.borsanyi@thphys.uni-heidelberg.de}\\[0.2cm]
{Universit\"at Heidelberg, Institut f\"ur
Theoretische Physik}\\
{Philosophenweg 16, 69120 Heidelberg, Germany}
}

\date{}

\begin{titlepage}
\maketitle
\def\thepage{}          

\begin{abstract}
\noindent
Transport equations can be derived from quantum field theory
assuming a loss of information about the details of the initial
state and a gradient expansion. While the latter can be
systematically improved, the assumption about a memory loss
is in general not controlled by a small expansion parameter. 
We determine the range of validity of transport equations
for the example of a scalar $g^2 \Phi^4$ theory.  
We solve the nonequilibrium time evolution using the
three-loop 2PI effective action. The approximation includes off-shell
and memory effects and assumes no gradient expansion.    
This is compared to transport equations to lowest order (LO)
and beyond (NLO). We find that the earliest time for the validity of 
transport equations is set by the characteristic relaxation time scale 
$t_{\rm damp} = - 2\omega/\Sigma^{\rm (eq)}_\varrho$, where 
$-\Sigma^{\rm (eq)}_\varrho/2$ denotes the on-shell imaginary-part 
of the self-energy. This time scale agrees with the
characteristic time for partial memory loss, but is much shorter than
thermal equilibration times. For times larger than 
about $t_{\rm damp}$ the gradient expansion to NLO is found to  
describe the ``full'' results rather well for $g^2 \lesssim 1$.   
\end{abstract}
\end{titlepage}

\renewcommand{\thepage}{\arabic{page}}


\section{Introduction}
\label{sec:intro}

Important phenomena in high-energy physics related to collision 
experiments of heavy nuclei, early universe cosmology and other complex
many-body systems urge a quantitative understanding 
of nonequilibrium dynamics in quantum field theories. 
One of the most crucial aspects concerns the characteristic time
scales on which thermal equilibrium is approached. 
Much of the recent interest derives from observations in collision 
experiments of heavy nuclei at RHIC. The experiments seem to indicate 
early thermalization, whereas the present theoretical understanding of 
QCD suggests a significantly longer thermal equilibration 
time~\cite{Heinz:2004pj}. 

Most theoretical estimates concerning the important question of thermalization 
have been obtained from transport theory, or assuming in addition
a quasi-particle picture, from kinetic theory. 
Transport equations can be derived from quantum field theory
assuming a loss of information about the details of the initial
state and a gradient expansion. While the latter can be
systematically improved, the assumption about a memory loss
is in general not controlled by a small expansion parameter. 
The time for a (partial) loss of memory represents a minimum 
characteristic time after which transport descriptions may be 
applied. It is important to quantify this characteristic time
in quantum field theory. The question is whether transport or 
kinetic theory can be used to quantitatively describe the 
early-time behavior, which is necessary if their application 
to the problem of fast apparent thermalization is viable.

With the advent of new computational techniques a direct 
account of quantum field degrees of freedom becomes increasingly 
feasible. There has been important progress in our 
understanding of nonequilibrium quantum fields using 
suitable resummation techniques based on 2PI generating
functionals~\cite{Berges:2004yj}. 
They have been successfully used to describe far-from-equilibrium dynamics 
and subsequent thermalization in quantum field 
theory~\cite{Berges:2000ur,Berges:2001fi,Cooper:2002qd,Berges:2002wr,Juchem:2003bi,Arrizabalaga:2005tf}. 
While it is difficult to test analytic approximations for theories such 
as QCD, where strong interactions play an important role, the description 
of scalar and fermionic theories for not too strong couplings seems 
well under quantitative control by now.

In this work we determine the range of applicability of transport
equations for a scalar $g^2 \Phi^4$ theory. 
The latter has been frequently studied in the past using the
required 2PI loop- \cite{Cornwall:1974vz} or 2PI $1/N$-expansions 
beyond lowest order~\cite{Berges:2001fi,Aarts:2002dj}.\footnote{Here $N$ 
denotes the number of field components.} 
Here we employ a three-loop 2PI effective action in $3 + 1$ dimensions. 
It includes direct scatterings, off-shell and memory effects. Most
importantly in this context, it employs no derivative expansion or
quasi-particle assumption. We solve the nonequilibrium evolution
for a class of nonequilibrium initial conditions reminiscent of some aspects 
of the anisotropic initial stage in the central region of two colliding wave 
packets. 

We then derive the corresponding transport equations to lowest order (LO) 
and next-to-lowest order (NLO) following standard prescriptions. Both
LO and NLO are then compared to the ``full'' results as described in 
Sec.~\ref{sec:comparison}. It turns out that the NLO corrections are crucial
already for relatively small couplings, which made it necessary to
go beyond the LO study presented in Ref.~\cite{Berges:2005ai}. 
The LO expressions are typically employed to obtain kinetic or Boltzmann 
equations using in addition 
a quasi-particle approximation. We emphasize that our comparisons do not 
rely on any kind of quasi-particle ansatz. We concentrate on the 
applicability of transport theory based on a derivative expansion, which 
also provides the necessary condition for the applicability of kinetic
equations.

\section{Exact evolution equations}

We start by considering the exact nonequilibrium evolution equation
for the time-ordered two-point correlation function
\bea
G(x,y) &=&  \langle T \Phi(x) \Phi(y) \rangle \nonumber\\
       &\equiv& F(x,y) - \frac{i}{2} \rho(x,y)\, {\rm sign}(x^0 - y^0) \, ,
\label{eq:decomp}
\eea
i.e.\ the expectation value of the product of two
Heisenberg field operators $\Phi$ with time-ordering $T$.
Here $\langle \ldots \rangle$ includes the trace over the
density matrix describing the initial state~\cite{Berges:2004yj}. 
For simplicity, we will consider a scalar $g^2 \Phi^4$ field theory
in the symmetric regime where the field expectation
value vanishes, i.e.\ $\langle \Phi(x) \rangle = 0$.

In general, there are two independent two-point functions,
which can be associated to the commutator and anti-commutator
of two fields. Accordingly,
the second line in Eq.~(\ref{eq:decomp}) is a rewriting, which
expresses the time-ordered product using 
\bea
\rho(x,y) &=& i \langle [\Phi(x),\Phi(y)] \rangle \, ,
\label{eq:comrho}\\
F(x,y) &=& \frac{1}{2} \langle \{\Phi(x),\Phi(y)\} \rangle \, .
\label{eq:anticomF}
\eea
Here $\rho(x,y)$ denotes the spectral function and 
$F(x,y)$ the statistical two-point function. For the
real scalar theory these determine the imaginary part
and the real part of the two-point function (\ref{eq:decomp}), respectively.
While the spectral function encodes the
spectrum of the theory, the statistical 
propagator gives information about occupation numbers.
Loosely speaking, the decomposition (\ref{eq:decomp}) makes explicit
what states are available and how often they are occupied.  
For nonequilibrium, $F(x,y)$ and $\rho(x,y)$ are in general two independent 
real-valued two-point functions. A simplification 
occurs in vacuum or thermal equilibrium where $F(x,y)$ and
$\rho(x,y)$ are related by the fluctuation-dissipation 
relation~\cite{Berges:2002wr}. We note that the spectral function  
encodes the equal-time commutation relations:
\beq
\rho(x,y)|_{x^0=y^0} = 0 \quad, \quad 
\partial_{x^0}\rho(x,y)|_{x^0=y^0} = \delta(\bx-\by) \, .
\label{eq:bosecomrel}
\eeq
 
The exact evolution equations for the 
statistical and spectral functions read~(for a detailed 
derivation see e.g.~\cite{Berges:2004yj}):
\bea 
\left[ \square_x 
+ M^2(x) \right] F(x,y) &=& 
- \int_0^{x^0}\! {\rm d}z^0 
\int_{-\infty}^{\infty}\! {\rm d}^d z\,
\Sigma_{\rho}(x,z) F(z,y)
\nonumber\\
&& + \int_0^{y^0}\! {\rm d}z^0 
\int_{-\infty}^{\infty}\! {\rm d}^d z\, 
\Sigma_F(x,z) \rho(z,y)  \, ,
 \label{eq:exactevolF} \\[0.1cm]
\left[\square_x + M^2(x) 
\right] \rho(x,y) &=& 
- \int_{y^0}^{x^0}\! {\rm d}z^0 
\int_{-\infty}^{\infty}\! {\rm d}^d z\, 
\Sigma_{\rho}(x,z) \rho(z,y) \, .
\label{eq:exactevolrho}
\eea
These are causal equations with characteristic 
``memory'' integrals, which integrate over the time history of the
evolution starting at time $t_0=0$. Since they are exact they are equivalent 
to any kind of identity for the two-point functions such as 
Schwinger-Dyson/Kadanoff-Baym equations. 
Eqs.\ (\ref{eq:exactevolF}) and (\ref{eq:exactevolrho})
are obtained for Gaussian initial conditions, which are also underlying 
transport equations (cf.\ Sec.~\ref{sec:transport}). 
The restriction to a class of initial conditions itself represents no 
approximation for the dynamics and for non-zero self-energies higher 
irreducible correlations build up for times $t > 0$.\footnote{More involved 
initial conditions can be considered using higher $n$-PI effective actions,
which would involve evolution equations for higher 
$n$-point functions beyond the two-point 
correlator~\cite{Berges:2004yj,Berges:2004pu,4PI}.}

The spectral ($\Sigma_\rho$) and statistical self-energy parts ($\Sigma_F$)
are obtained from the proper self-energy $\Sigma(x,y)$, which sums all 
one-particle irreducible diagrams, using a decomposition similar to
Eq.~(\ref{eq:decomp})~\cite{Berges:2001fi}:
\bea
\Sigma(x,y) &=& - i \Sigma^{(0)}(x) \delta(x-y)
\nonumber\\
&& + \Sigma_F(x,y) - \frac{i}{2} \Sigma_{\rho}(x,y)\, 
{\rm sign}(x^0 - y^0) \, . 
\label{eq:sighominh}
\eea
Since $\Sigma^{(0)}$ corresponds to a space-time dependent 
mass-shift it is convenient to introduce the notation
\beq
M^2(x;G) = m^2 + \Sigma^{(0)}(x;G) \, .
\label{eq:localself}
\eeq
To close the set of equations of motion
(\ref{eq:exactevolF}) and (\ref{eq:exactevolrho}) we need further equations
that connect the self-energies to $F$ and $\rho$. This
link is established by defining an approximation scheme based on the
two-particle irreducible (2PI) effective 
action~\cite{Cornwall:1974vz},\footnote{For solving 
Eqs.~(\ref{eq:exactevolF})--(\ref{eq:exactevolrho}) we 
introduce a further approximation by restricting
the memory time integration to a maximum interval in order to 
fit in the memory of the computing device. In this paper we use 
a $8192^2\times16^3$ lattice
using about 100 GB on a distributed memory super-computer. The used integration
interval is $\sim 1200 m_R^{-1}$ for $g^2=0.5$ and $\sim 1700 m_R^{-1}$ for
$g^2=1$. This interval is much longer than the characteristic memory or 
damping time of the propagator, thus this memory-cut has a negligible effect on
the dynamics.} which to three-loop order in the $g^2\Phi^4$ scalar theory 
yields the self-energies
\bea
\Sigma_F(x,y)&=&-96g^4F(s,\vec x)\left(F^2(x,y)
-\frac34\rho^2(x,y)\right) \, ,
\label{eq:SigmaF}\\
\Sigma_\rho(x,y)&=&-288g^4\rho(x,y)\left(F^2(x,y)
-\frac1{12}\rho^2(x,y)\right) \, .
\label{eq:Sigmarho}
\eea
This approximation has been shown to describe 
thermalization~\cite{Berges:2000ur,Juchem:2003bi,Arrizabalaga:2005tf} 
and includes direct scatterings, off-shell and memory effects. Therefore,
it goes far beyond standard kinetic descriptions or Boltzmann equations.
Most importantly in this context, it employs no derivative expansion.
The latter is a basic ingredient for transport or kinetic theory.

\section{Transport equations}
\label{sec:transport}

The spectral function (\ref{eq:comrho}) is directly related to the
retarded propagator, $G_R$, or the advanced one, $G_A$, by
\beq
G_R(x,y)=\Theta(x^0-y^0)\rho(x,y)\, , \quad
G_A(x,y)=-\Theta(y^0-x^0)\rho(x,y)
\label{eq:GRGA}\, .
\eeq
Similarly, the retarded and advanced self-energies are 
\beq
\Sigma_R(x,y)=\Theta(x^0-y^0)\Sigma_\rho(x,y)\, , \quad
\Sigma_A(x,y)=-\Theta(y^0-x^0)\Sigma_\rho(x,y)
\label{eq:SRSA}\, .
\eeq
With the help of this notation,
interchanging $x$ and $y$ in the evolution equation
(\ref{eq:exactevolF}) and subtraction one obtains
\begin{eqnarray}
\lefteqn{\Big( \square_x - \square_y + M^2 \left( x \right) 
- M^2 \left( y \right) \Big) F \left( x, y \right)}  
\nonumber \\
  & = & \int {\rm d}^{d+1}{z}\,  \theta \left( z^0 \right) 
\Big( F \left( x, z \right) \Sigma_A \left( z, y \right) 
+ G_R \left( x, z \right) \Sigma_F \left( z, y \right) 
\nonumber \\ 
  &   & {} - \Sigma_R \left( x , z \right) F \left( z, y \right) 
- \Sigma_F \left( x, z \right) G_A \left( z, y \right) \Big) 
\label{eq:Fint} \, .
\end{eqnarray}
The same procedure yields for the spectral function
\begin{eqnarray}
\lefteqn{\Big( \square_x - \square_y + M^2 \left( x \right) 
- M^2 \left( y \right) \Big) \rho \left( x, y \right)} 
\nonumber \\
  & = & \int {\rm d}^{d+1}{z}\, 
\Big( G_R \left( x, z \right) \Sigma_{\rho} \left( z, y \right) 
+ \rho \left( x, z \right) \Sigma_A \left( z, y \right) 
\nonumber \\ 
  &   & {} - \Sigma_{\rho} \left( x , z \right) G_A \left( z, y \right) 
- \Sigma_R \left( x, z \right) \rho \left( z, y \right) \Big) 
\label{eq:rhoint} \,.
\end{eqnarray}
So far, the equations (\ref{eq:Fint}) and (\ref{eq:rhoint}) 
are fully equivalent to the exact equations (\ref{eq:exactevolF})
and (\ref{eq:exactevolrho}).

\subsection{Assumptions}
\label{sec:assumptions}

Transport equations are obtained from the exact equations 
(\ref{eq:Fint}) and (\ref{eq:rhoint}) by the
following prescription:
\begin{enumerate}
\item[1)] 
Drop the $\theta$-function in Eq.~(\ref{eq:Fint}), which
sets the time $t_0=0$ where the system is initialized. 
Dropping it amounts to sending the initial time to the infinite past,
i.e.\ $t_0 \to -\infty$. 
Of course, a system that thermalizes would have reached already
equilibrium at any finite time if initialized in the remote past. 
Therefore, in practice a ``hybrid'' description is employed: 
transport equations are
initialized by prescribing $F$, $\rho$ and derivatives 
at a {\em finite} time using the equations with $t_0 \to -\infty$ 
as an approximate description. 
\item[2)] Employ a gradient expansion to Eqs.~(\ref{eq:Fint}) and
(\ref{eq:rhoint}).  In practice, this expansion is carried out to lowest order
(LO), i.e.\ no derivatives, or next-to-lowest order (NLO) in the number of
derivatives with respect to the center coordinates
\beq
X^{\mu} \equiv \frac{x^{\mu} + y^{\mu}}{2} 
\label{eq:center}
\eeq
and powers of the relative coordinates
\beq
s^{\mu} \equiv x^{\mu} - y^{\mu} \, .
\label{eq:relative}
\eeq
\item[3)] Even for finite $X^0$ one assumes that the 
relative-time coordinate $s^0$ ranges from $-\infty$ to $\infty$
in order to achieve a convenient description in Wigner space, i.e.\ in Fourier 
space with respect to the relative coordinates (\ref{eq:relative}).

\end{enumerate} 
We emphasize that the {\em ad hoc} approximations 1) and 3)
are in general not controlled by a small expansion parameter.
They require a loss of information about the details of the initial state.
More precisely, they can only be expected to be valid for sufficiently
late times times $t$ when initial-time correlations   
become negligible, i.e.\ $\langle \Phi(0,\bx) \Phi(t,\by) \rangle \simeq 0$. 
The standard approximations 1) and 3) may, in
principle, be evaded. However, if they are not applied then 
a gradient expansion would become too cumbersome to be of use 
in practical calculations. 

The derivation of transport equations has been discussed 
in great detail in the
literature~\cite{gradient,Danielewicz:1982kk,Calzetta:1986cq,Boyanovsky:1996xx,Wong:1996ta,Ivanov:1998nv,LMS,Leupold:1999ga,Blaizot:2001nr,Jakovac:2002rc,Berges:2002wt,Prokopec:2003pj,Ikeda:2004in}.
Most discussions focus on kinetic theory employing the additional 
approximation of a suitable quasi-particle ansatz, which goes beyond 
assumptions 1) -- 3). In order to be general, we do not restrict the
following analysis to any quasi-particle picture.

\begin{figure}[t]
\centerline{
\epsfig{file=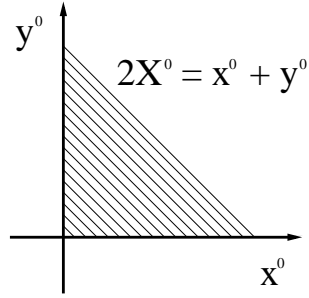,width=4.cm}
}
\caption{\small 
For given finite $2X^0 = x^0 + y^0$ the relative coordinate 
$s^0 = x^0 - y^0$ has a finite range from $-2 X^0$ to $2 X^0$.
\label{fig:srange}
}
\end{figure}

For the description in Wigner space 
we introduce the Fourier transforms with respect to
the relative coordinates, such as
\bea
\tilde{F}(X;\omega,\bp)&=&\int_{-2X^0}^{2X^0}\! {\rm d}s^0\, e^{i\omega s^0} 
\int_{-\infty}^{\infty}\! {\rm d}^d s\, e^{i \bp\, \bs} F(X+s/2,X-s/2)
\nonumber\\
&=& 2 \int_{0}^{2X^0} \!{\rm d}s^0\, \cos(\omega s^0)  
\int_{-\infty}^{\infty}\! {\rm d}^d s\, e^{i \bp\, \bs} F(X+s/2,X-s/2)
\label{eq:Frho}, \qquad 
\eea
using the symmetry property 
$F(x,y) = F(y,x)$ for the second line.
We emphasize that the time integral
over $s^0$ is bounded by $\pm 2 X^0$. The time
evolution equations are initialized at time
$x^0 = y^0 = 0$ such that $x^0 \ge 0$ and $y^0 \ge 0$. 
According to Eq.~(\ref{eq:relative}),
the minimum value of the relative coordinate $s^0$ is then given by
$-y^0 = - 2X^0$ for $x^0 = 0$ while its maximum value is $x^0 = 2X^0$ 
for $y^0 =0$ as illustrated in Fig.~\ref{fig:srange}. Similarly, we define 
\bea
\tilde{\varrho}(X;\omega,\bp)&=&
-i\int_{-2X^0}^{2X^0}\! {\rm d}s^0\, e^{i\omega s^0} 
\int_{-\infty}^{\infty}\! {\rm d}^d s\, e^{i \bp\, \bs}
\rho(X+s/2,X-s/2)
\nonumber\\
&=& 2 \int_{0}^{2X^0} \! {\rm d}s^0\, 
\sin(\omega s^0) \int_{-\infty}^{\infty}\! {\rm d}^d s\, e^{i \bp\, \bs}
\rho(X+s/2,X-s/2), \qquad 
\label{eq:Wigrho}
\eea
where a factor of $i$ is included in the definition 
to have $\tilde{\varrho}$ real and we 
have used $\rho(x,y) = - \rho(y,x)$. It will be convenient to also 
introduce the cosine transform
\bea
\tilde{\varrho}_+(X;\omega,\bp)&=&
2 \int_{0}^{2 X^0} \!\!\! {\rm d}s^0\, 
\cos(\omega s^0) \int_{-\infty}^{\infty}\! {\rm d}^d s\, e^{i \bp\, \bs}
\rho(X+s/2,X-s/2) . \qquad
\label{eq:Wigrhoplus}
\eea
The equivalent transformations
are done to obtain the self-energies $\tilde{\Sigma}_F(X;\omega,\bp)$,
$\tilde{\Sigma}_\varrho(X;\omega,\bp)$ and the cosine transform
$\tilde{\Sigma}_+(X;\omega,\bp)$ corresponding to Eq.~(\ref{eq:Wigrhoplus}).
From Eqs.~(\ref{eq:GRGA}) and (\ref{eq:SRSA}) it follows that the 
cosine transforms are related to the real part of the retarded
functions
\beq
\tilde{\varrho}_+(X;\omega,\bp) \,\equiv\, 2\, 
{\rm Re} \tilde{G}_R(X;\omega,\bp)
\,\, , \quad
\tilde{\Sigma}_+(X;\omega,\bp) \,\equiv\, 2\, 
{\rm Re} \tilde{\Sigma}_R(X;\omega,\bp)
\, .
\eeq

In order to exploit the convenient properties of a
Fourier transform it is a standard procedure to extend
the limits of the integrals
over the relative time coordinate
in Eqs.~(\ref{eq:Frho}), (\ref{eq:Wigrho}) and (\ref{eq:Wigrhoplus})
to $\pm\infty$. For instance, using the chain rule 
\beq 
\square_x - \square_y = 2 \partial_{s_{\mu}} 
\partial_{X^{\mu}} 
\eeq 
and the gradient expansion of the mass terms in Eq.~(\ref{eq:Fint})
to\footnote{The expression is actually correct to NNLO
since the first correction is 
${\cal O}\left(\left(s^{\mu}\partial_{X^{\mu}}\right)^3\right)$.}
NLO:
\beq 
M^2 \left( X + \frac{s}{2} \right) - M^2 \left( X - \frac{s}{2} \right) 
\,\simeq\, s^{\mu} \partial_{X^{\mu}} M^2 \left( X \right) 
\,,
\eeq
then the LHS of Eq.~(\ref{eq:Fint}) becomes in Wigner space
\bea
\int_{-\infty}^{\infty} {\rm d}^{d+1} s\, e^{i p s} 
\left[ 2 \partial_{s_{\mu}} \partial_{X^{\mu}} 
+ s^{\mu} \partial_{X^{\mu}} M^2 \left( X \right) 
\right] 
F \left( X+s/2, X-s/2 \right) &&
\nonumber\\
   =  - i \left[ 2 p^{\mu} \partial_{X^{\mu}} 
+  \left( \partial_{X^{\mu}} M^2 \left( X \right) \right)
\partial_{p_{\mu}} \right] \tilde{F} \left( X; p \right)\, ,&& 
\eea
with $p^0 \equiv \omega$. Similarly, one transforms the
RHS of Eqs.~(\ref{eq:Fint}) and (\ref{eq:rhoint}) using
the above prescription 1) -- 3).

\subsection{LO equations}

The LO equations are obtained by 
neglecting ${\cal O}\left(\partial_{X^{\mu}} 
\partial_{p_{\mu}}\right)$
and higher contributions in the gradient expansion. To this
order the transport equations then read:
\bea
2 p^{\mu} \partial_{X^{\mu}} \tilde{F} \left( X; p \right)
&=&
\tilde{\Sigma}_{\varrho} \left( X; p \right) 
\tilde{F} \left( X; p \right) 
- 
\tilde{\Sigma}_F \left( X; p \right)
\tilde{\varrho} \left( X; p \right) 
\, ,
\label{eq:LOgradF}\\[0.2cm]
2 p^{\mu} \partial_{X^{\mu}} \tilde{\varrho} \left( X; p \right)
&=& 0 \, .
\label{eq:LOgradrho}
\eea
We note that the compact form of the gradient expanded equation
to LO is very similar to the exact equation for the statistical
function (\ref{eq:exactevolF}). The main technical difference is that 
there are no integrals over the time history. Furthermore, to this
order in the expansion the evolution equation for the
spectral function $\tilde\varrho(X;p)$ becomes trivial.
Accordingly, this approximation describes changes in the
occupation number while neglecting its impact on the
spectrum and vice versa.
The LO equations provide the basis of almost all practical 
applications of transport and Boltzmann or kinetic equations.

\subsection{NLO equations}

To NLO, i.e.\ neglecting all ${\cal O}\left(\left(\partial_{X^{\mu}} 
\partial_{p_{\mu}}\right)^2\right)$ and higher contributions,
the transport equations become substantially more involved:
\bea
\lefteqn{\left[ 2 p^{\mu} \partial_{X^{\mu}} 
+ \left( \partial_{X^{\mu}} M^2 \left( X \right) \right) 
\partial_{p_{\mu}} \right] \tilde{F} \left( X; p \right)} 
\nonumber \\
& = & 
\tilde{\Sigma}_{\varrho}\left( X; p \right)
\tilde{F} \left( X; p \right) 
-\tilde{\Sigma}_F \left( X; p \right) 
 \tilde{\varrho} \left( X; p \right) 
\nonumber \\
& + &  \frac{1}{2}\left\{ \tilde{\Sigma}_F 
\left( X; p \right) , \tilde{\varrho}_+\left( X; p \right) 
\right\}_{PB} + \frac{1}{2} \left\{ \tilde{\Sigma}_+ 
\left( X; p \right) , \tilde{F} \left( X; p \right) 
\right\}_{PB} \, ,
\label{eq:NLOF}\\[0.2cm]
\lefteqn{\left[ 2 p^{\mu} \partial_{X^{\mu}} 
+ \left( \partial_{X^{\mu}} M^2 \left( X \right) \right) 
\partial_{p_{\mu}} \right] \tilde{\varrho} \left( X; p \right)} 
\nonumber \\
& = & \frac{1}{2}\left\{ \tilde{\Sigma}_{\varrho} \left( X; p \right) , 
\tilde{\varrho}_+ \left( X; p \right) \right\}_{PB} 
+ \frac{1}{2}\left\{ \tilde{\Sigma}_+ \left( X; p \right) , 
\tilde{\varrho} \left( X; p \right) \right\}_{PB} 
\label{eq:NLOrho} \, .
\eea
Here we have introduced the Poisson brackets
\bea
\lefteqn{\left\{ \tilde{f} \left( X; p \right) ; \tilde{g} 
\left( X; p \right) \right\}_{PB}}
\nonumber\\
& \equiv & 
\left( \partial_{p_{\mu}} \tilde{f} \left( X; p \right) 
\right) \left( \partial_{X^{\mu}} \tilde{g} \left( X; p \right) \right) 
- \left( \partial_{X^{\mu}} \tilde{f} \left( X; p \right) \right) 
\left( \partial_{p_{\mu}} \tilde{g} \left( X; p \right) \right) \, .
\label{eq:poisson}
\eea
The somewhat tedious NLO expressions are more or less not applied in
practice. The rapidly increasing complexity of transport equations
at even higher order seems forbiddingly difficult. In this case one may
better employ the comparably compact full expressions without 
gradient expansion anyways.

\section{Comparison of LO, NLO and ``full'' results}
\label{sec:comparison}

Nonequilibrium dynamics requires the specification
of an initial state. While the corresponding initial conditions 
for the spectral function are
governed by the commutation relations (\ref{eq:bosecomrel}),
the statistical function $F(x,y)$ and first derivatives  
at $x^0 = y^0 = 0$ have to be specified. 
Following Ref.~\cite{Berges:2005ai}, we consider a spatially homogeneous 
situation with an initially high occupation number 
of modes moving in a narrow momentum range around  
$p_3 \equiv p_\pa = \pm p_{\rm ts}$ along the ``beam direction''. 
The occupation
numbers for modes with momenta perpendicular to this direction, 
$p_1^2 + p_2^2 \equiv p_{\pe}^2$, are small or vanishing.
The situation is reminiscent of some aspects of the anisotropic
initial stage in the central region of two colliding wave 
packets.\footnote{Other interesting scenarios
include ``color-glass''-type initial conditions with 
distributions $\sim \exp(-\sqrt{p_\pe^2}/Q_s)$ peaked around $p_3 =0$
with ``saturation'' momentum~$Q_s$.} Of course,
a peaked initial particle number distribution is not
very specific and is thought  
to exhibit characteristic properties
of nonequilibrium dynamics for a large variety
of physical situations.
  
\begin{figure}[t]
\centerline{
\epsfig{file=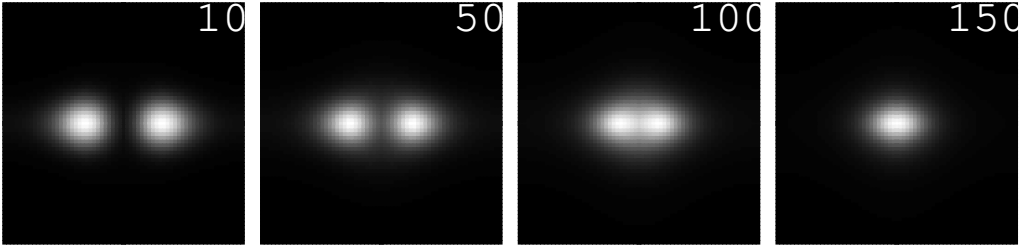,width=\textwidth}
}
\caption{\small Snapshots at times 
$X^0\, m_R =10$, $50$, $100$ and $150$ of the 
initially anisotropic occupation number distribution as 
a function of $p_{\pe}$ (vertical) and 
$p_{\pa}$ (horizontal). Bright (dark) regions correspond
to high (low) occupation numbers. The shown resolution
was achieved by using a $64^3$ spatial lattice.
\label{fig:snapshots}
}
\end{figure}

The initial distributions do not correspond to thermal equilibrium,
which is isotropic in momenta. Interactions will lead to the
isotropization of the distribution, which finally approaches thermal
equilibrium.     
An example of the earlier stages of such an evolution is shown 
in Fig.~\ref{fig:snapshots} 
at various times in units of the renormalized thermal mass
$m_R$. Shown is the distribution as a function of $p_{\pe}$ (vertical) and 
$p_{\pa}$ (horizontal), where bright (dark) regions correspond
to high (low) occupation numbers. One observes that
for times exceeding some characteristic isotropization 
time of about $\simeq 100/m_R$ 
the distribution starts to become rather
independent of the momentum direction. 
For a time of $150/m_R$ the figure shows already an almost perfectly 
isotropic situation. It has been
shown in Ref.~\cite{Berges:2005ai} that the
characteristic time scale for isotropization
is well described by the relaxation or damping time  
\beq
t_{\rm damp} \,=\, 
- \frac{2 \omega^{(\rm eq)}}{\tilde{\Sigma}^{(\rm eq)}_{\varrho}}\,,
\label{eq:estimate}
\eeq
where the imaginary part of the thermal equilibrium 
self-energy $-\Sigma^{(\rm eq)}_{\varrho}/2$
is evaluated for on-shell frequency $\omega^{(\rm eq)}$ for 
momentum $p_{\rm ts}$. This time scale agrees with
the characteristic time for the effective loss of details about
the initial conditions~\cite{Berges:2001fi,Berges:2002wr}.
The time $t_{\rm damp}$ for partial memory loss is an important 
characteristic time scale
in nonequilibrium dynamics, which is very different from the
thermal equilibration time. The distribution is still
far from equilibrium and the approach
to a thermal Bose-Einstein distribution
takes substantially longer than $t_{\rm damp}$.  
For the employed initial conditions and coupling strength the 
characteristic thermal equilibration times 
exceed $t_{\rm damp}$ by more than two orders of 
magnitude~\cite{Berges:2005ai}. 

More precisely, we investigate a class of spatially homogeneous 
initial conditions parametrized as 
\beq
F(x^0,y^0;\bp)|_{x^0=y^0=0} 
= \frac{n_0(\bp)+1/2}{\omega_p} \, ,
\label{eq:init1}
\eeq
with $\partial_{x^0}F(x^0,0;\bp)|_{x^0=0} = 0$,  
$\partial_{x^0}\partial_{y^0}F(x^0,y^0;\bp)|_{x^0=y^0=0} = [n_0(\bp)+1/2]\, 
\omega_p$
for $\omega_p \equiv \sqrt{p_{\pe}^2 + p_{\pa}^2 + M_0^2}$ and $M^2_0 
\equiv M^2(x^0=0)$. The initial distribution function $n_0(\bp)$
is peaked around the ``tsunami'' momentum $p_{ts}$ with amplitude
$A$ and width $\sigma$:
\beq
n_0(\bp) = A\, \exp\left\{-\frac{1}{2\sigma^2}
\left[p_{\pe}^2 + (|p_{\pa}| - p_{\rm ts})^2 
\right]\right\} \, .
\label{eq:init2}
\eeq
We compute the nonequilibrium dynamics for a rather weak coupling of the 
$g^2 \Phi^4$--interaction. 
For Fig.~\ref{fig:snapshots} we use $g^2=0.25$ and an initial distribution
peaked at $p_{\rm ts}=0.8\,m_R$ with $A=20$ and $\sigma = 0.32\,m_R$. 
For these parameters we obtain $t_{\rm damp} \simeq 100/m_R$ according to
Eq.~(\ref{eq:estimate}). We will consider a range of different couplings and 
amplitudes below. 

The question is whether transport or kinetic equations can be used 
to quantitatively describe the early-time behavior before $t_{\rm damp}$, 
which is necessary if their application to the problem of fast 
apparent thermalization explained in Sec.~\ref{sec:intro} is viable. 
This is discussed in the following comparing to LO or NLO transport 
equations.

\subsection{Isotropization}
\label{sec:iso}
 
\begin{figure}[t]
\begin{center}
\epsfig{file=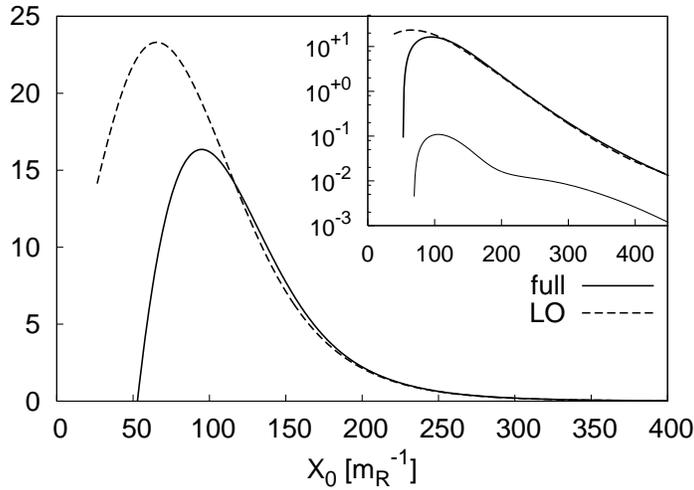,width=10.cm}
\end{center}
\vspace*{-0.6cm}
\caption{\small 
The solid line shows the on-shell 
$\partial_{X^0} \Delta \tilde{F}$ in Wigner 
coordinates, as computed from 
the solution of Eqs.~(\ref{eq:exactevolF})--(\ref{eq:exactevolrho})
without a gradient
expansion. The dashed line represents the same quantity
using the expression from the LO gradient expansion 
according to Eq.~(\ref{eq:LOgradF}).
If the gradient expansion to lowest order is correct, then
both lines have to agree. 
For this quantity one observes that the lowest-order gradient expansion
becomes valid on a time scale of about the 
isotropization or damping time $t_{\rm damp} \simeq 100/m_R$.
The inset shows the same results on a 
logarithmic scale to make the small on-shell
$\partial_{X^0} \Delta \tilde{\varrho}$ visible.
Its full result is given by the solid lower line,
whereas the LO contribution is zero.
} 
\label{fig:wignerrel}
\end{figure}
A characteristic anisotropy measure can be chosen as~\cite{Berges:2005ai}
\beq
\Delta \tilde{F}(X^0;\omega,p_{\rm ts}) \equiv 
\tilde{F}(X^0;\omega,\bp)|_{p_{\pe}=0,p_{\pa}=p_{\rm ts}}
- \tilde{F}(X^0;\omega,\bp)|_{p_{\pe}=p_{\rm ts},p_{\pa}=0} \, ,
\label{eq:anisoF}
\eeq
which vanishes for the case of an isotropic correlator since the
latter depends on $\bp^2 = p_\pe^2+p_\pa^2$. 
Fig.~\ref{fig:wignerrel} shows the time evolution of the on-shell 
derivative $ \partial_{X^0} 
\Delta \tilde{F}(X^0;\omega(p_{\rm ts}),p_{\rm ts})$~\cite{Berges:2005ai}.
The thick curve represents the full result, which 
is obtained from solving the evolution equations
(\ref{eq:exactevolF}) and (\ref{eq:exactevolrho}) for the
self-energies (\ref{eq:SigmaF}) and (\ref{eq:Sigmarho}) without
any gradient expansion.
For comparison, we evaluate the same quantity
using the LO gradient expansion according to Eq.~(\ref{eq:LOgradF}). 
For this we evaluate the RHS of Eq.~(\ref{eq:LOgradF}) using the full result  
for $\tilde{F}(X^0;\omega,\bp)$ and $\tilde{\varrho}(X^0;\omega,\bp)$.
If the gradient expansion to lowest order 
is correct, then both results have to agree. Indeed, one observes from
Fig.~\ref{fig:wignerrel}, employing the same parameters as for
Fig.~\ref{fig:snapshots}, that
the curves for the LO (dashed) and the full result agree well at
sufficiently late times. However, they only agree {\em after} some time,
which is rather well determined by
the characteristic time scale $\sim t_{\rm damp}$ for the effective 
loss of details about the initial conditions. 
The latter is given by the thermal equilibrium estimate
(\ref{eq:estimate}). Similarly, one can observe this time
scale from the decay of the unequal-time spectral function in
thermal equilibrium, $\rho^{\rm (eq)}(t;\bp_{\rm ts})$
as explained below.
The latter measures the characteristic 
decay of correlations at time $t$ with the
initial state. The decay time for $\rho^{\rm (eq)}$
coincides rather well with the time for the onset of validity 
of the LO result. 
\begin{figure}[t]
\centerline{
\epsfig{file=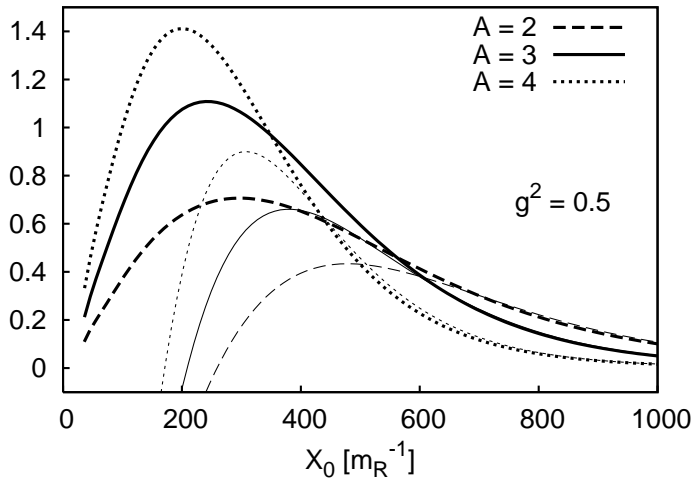,width=10.cm}
\vspace*{-0.5cm}
}
\caption{\small 
Same as in Fig.~\ref{fig:wignerrel} but for amplitudes $A = 2,3,4$,
i.e.\ smaller occupation numbers according to Eq.~(\ref{eq:init2}), 
and coupling $g^2 = 0.5$. The lower solid/dashed/dotted curves
represent the full result, while the upper (thick) ones show the respective
LO approximation.  
}
\label{fig:variousA}
\end{figure}

\begin{figure}[t]
\centerline{
\epsfig{file=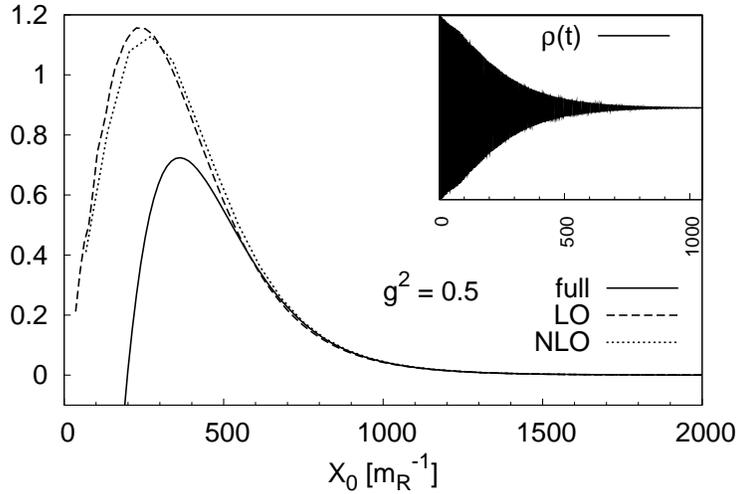,width=10.cm}
\vspace*{-0.5cm}
}
\caption{\small 
Comparison of LO, NLO and full result for the on-shell 
$\partial_{X^0} \Delta \tilde{F}$
with $A=3$ and $g^2 = 0.5$. For this quantity 
LO and NLO results do not differ much, while
there is a substantial deviation from the full result at early times.
The inset displays the real-time equilibrium
spectral function for the considered mode, which  
exhibits the characteristic damping time for (partial) memory loss.
The latter agrees rather well with the time for the onset of validity
of a transport description.}
\label{fig:LONLO}
\end{figure}

Using smaller amplitudes $A$ (with $\sigma$ and $p_{\rm ts}$ kept fixed) 
one observes a similar picture. In Fig.~\ref{fig:variousA}
we compare runs for $A=2$, $3$ and $4$ with $g^2=0.5$. For instance, 
for $A=3$ the LO order and the full result approach each other rather
closely after a time of about $500/m_R$. This time is rather well 
described by the characteristic decay time of the unequal-time
correlator $\rho^{\rm (eq)}$ as shown in the inset of 
Fig.~\ref{fig:LONLO}. Of course, the precise notion
of a time after which a transport description holds depends
on the definition and prescribed accuracy. We find for
$A=2,3,4$ that the damping time according to
Eq.~(\ref{eq:estimate}) is  $t_{\rm damp} = 270, 183, 141 m_R^{-1}$.
For $A=3$ this time may be compared to the inset of 
Fig.~\ref{fig:LONLO}, which shows that only after, say,
a time of about $500 m_R^{-1}$ the correlator is small
enough that initial-time effects play no important role.

One may expect that the failure of the LO transport equations
to describe the dynamics before isotropization completes 
is due to substantial NLO contributions. However, this turns
out not to be the case. Fig.~\ref{fig:LONLO}
compares LO, NLO and full result for the time evolution of the 
on-shell $\partial_{X^0} \Delta \tilde{F}(X^0;\omega(p_{\rm ts}),p_{\rm ts})$
for $A=3$ and $g^2 = 0.5$.
The solid line represents the result as computed from 
the solution of Eq.~(\ref{eq:exactevolF}) without 
a gradient expansion. As before, the dashed line shows the same quantity 
according to the LO expression given by the RHS of Eq.~(\ref{eq:LOgradF})
using the full results $\tilde{F}(X^0;\omega,\bp)$ and
$\tilde{\varrho}(X^0;\omega,\bp)$. Similarly, we evaluate 
the RHS of the NLO expression given by the RHS of 
Eq.~(\ref{eq:NLOF}).\footnote{More precisely, we employ the
Poisson-Brackets (\ref{eq:poisson}) where the   
$\partial_{X}$-derivatives are calculated from the LO
expressions (\ref{eq:LOgradF}) and (\ref{eq:LOgradrho}), 
which is correct to NLO.} The result corresponds to the dotted
line in Fig.~\ref{fig:LONLO}, which turns out to be rather close
to the LO curve.  

\begin{figure}[t]
\centerline{
\epsfig{file=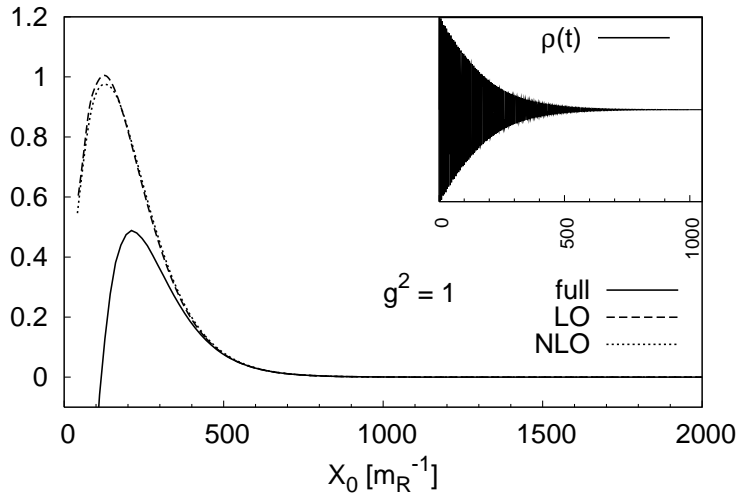,width=10.cm}
\vspace*{-0.6cm}
}
\caption{\small 
Same as in Fig.~\ref{fig:LONLO} but for $g^2 = 1$.
Even for this stronger coupling the NLO corrections remain small, while
there is a substantial deviation from the full result at early times.  }
\label{fig:one}
\end{figure}

For the quantity $\partial_{X^0} 
\Delta \tilde{F}(X^0;\omega(p_{\rm ts}),p_{\rm ts})$ the NLO 
corrections remain small even for rather strong couplings.
This is demonstrated in Fig.~\ref{fig:one} for $g^2 = 1$ 
and same initial conditions as for Fig.~\ref{fig:LONLO}. 
The convergence of the gradient expansion itself is 
apparently quite good here even for couplings of order one.  
However, when studying the importance of NLO corrections one 
has to distinguish
between quantities whose vanishing corresponds to an enhanced
symmetry from more generic quantities. Once an ensemble is
isotropic it always remains so at later times even though the
dynamics may still be far from equilibrium. The vanishing of
$\Delta \tilde{F}$ obviously corresponds to an enhanced symmetry,
i.e.\ rotation symmetry. In the 
following we consider more generic cases. 

\subsection{Thermal equilibration}
\label{sec:thermal}

It has been demonstrated in Ref.~\cite{Berges:2005ai}
that in a scalar theory
isotropization happens much faster than the approach to thermal 
equilibrium. Only a subclass
of those processes that lead to thermalization are actually
required for isotropization: ``ordinary''
$2 \leftrightarrow 2$ scattering processes are sufficient in
order to isotropize a system, while global particle number 
changing processes are crucial to approach thermal 
equilibrium. 

\begin{figure}[t]
\centerline{
\epsfig{file=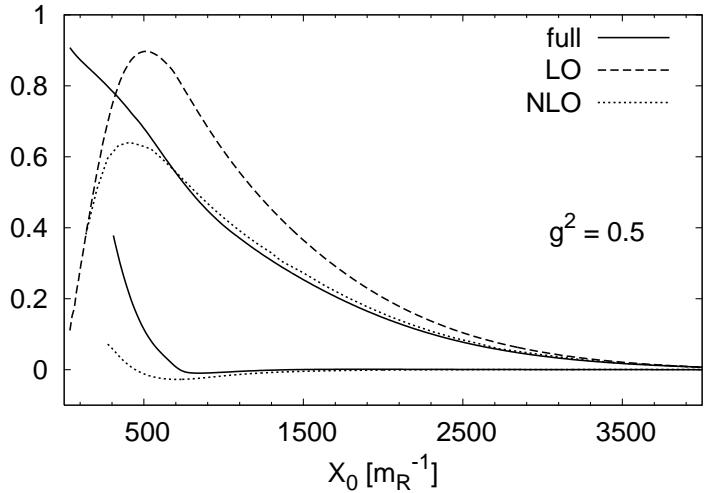,width=10.cm}
\vspace*{-0.6cm}
}
\caption{\small 
Comparison of LO, NLO and full result for the on-shell
$\partial_{X^0} \tilde{F}$ at zero momentum
with $A=3$ and $g^2 = 0.5$. The lower curves show the full and
NLO result for $\partial_{X^0}\tilde\varrho$,
to lowest order there is no contribution.
The LO transport equation fails to
describe the full results until rather late times. In contrast, taking into
account the substantial NLO corrections the gradient expansion
becomes quite accurate for times larger than about the
characteristic damping time for partial memory loss.
}
\label{fig:thermal0.5}
\end{figure}

Above we have investigated a quantity measuring an anisotropy
($\partial_{X^0}\Delta \tilde{F}$), and which, therefore,
approaches zero rather quickly after $t_{\rm damp}$. 
Consequently, it is not affected by
the relevant late-time processes which finally lead to a Bose-Einstein
distribution. The accuracy of the gradient expansion
for the late-time behavior can be investigated considering 
$\partial_{X^0} \tilde{F}(X^0;\omega,\bp)$ instead,
which does not vanish for an isotropic ensemble.  Fig.~\ref{fig:thermal0.5}
shows the time evolution for this quantity on shell at zero momentum for the
same parameters as for Fig.~\ref{fig:LONLO}. The solid 
lines give the full result for $ \partial_{X^0} \tilde{F}$ 
(upper curve) and for $\partial_{X^0} \tilde{\varrho}$ (lower curve). 
This is compared to the LO (dashed) and NLO (dotted) results evaluated 
in the same way as explained above. 

\begin{figure}[t]
\centerline{
\epsfig{file=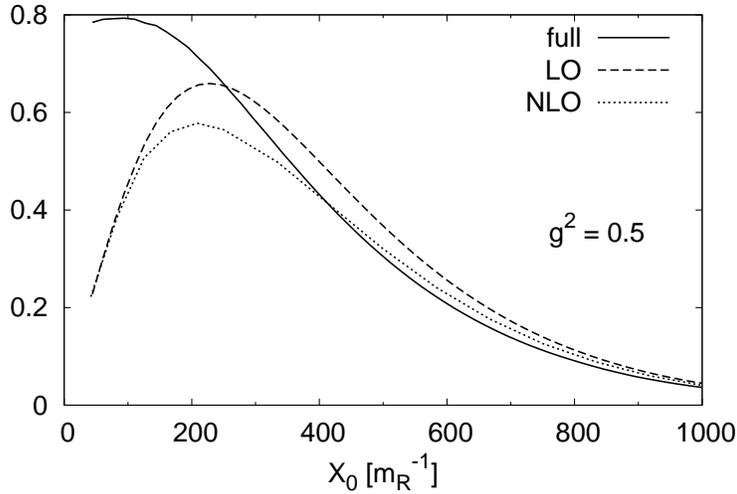,width=10.cm}
\vspace*{-0.6cm}
}
\caption{\small 
Same as in Fig.~\ref{fig:thermal0.5} but for $g^2 = 1$.}
\label{fig:thermal1}
\end{figure}

One observes from Fig.~\ref{fig:thermal0.5} that the LO and the full
result approach each other only at a time much later than 
it happens for the anisotropy quantity $\Delta \tilde{F}$ discussed
in Sec.~\ref{sec:iso}. However, one
also observes from Fig.~\ref{fig:thermal0.5} that here the NLO corrections
are substantial. The NLO results are much closer to
the full result for times larger than about $\sim t_{\rm damp}$.
We emphasize that the employed coupling $g^2 = 0.5$ is not very large.
The situation is qualitatively not very different also for couplings
of order one, which is exemplified in Fig.~\ref{fig:thermal1} for
the same parameters as for Fig.~\ref{fig:one}. One observes, however, 
that the deviations of the NLO from the full result are stronger
even at rather late times in this case. 

We conclude that quantities which are sensitive to the late-time
dynamics can receive important corrections at NLO in the gradient 
expansion. This finding is, of course, not surprising for strong couplings.
However, sizeable NLO corrections can be obtained already for couplings
of about $g^2 \simeq 1/4$. Nevertheless, we observe that for not too early 
times the NLO result can get rather close to the full result for
$g^2 \lesssim 1$. The agreement of the (NLO) transport description and full 
results start at about the same time as for the isotropization case
discussed above. Times shorter than about $t_{\rm damp}$ are clearly
beyond the range of validity of transport equations
even for weak couplings. This limitation
is apparently not a consequence of a gradient expansion to low orders.
The latter seems to converge rather well for weak enough couplings. 

The failure of transport equations to describe early-time dynamics 
seems mainly based on the fact 
that the underlying assumptions 1) and 3) of Sec.~\ref{sec:assumptions}
are not controlled by a small parameter. The validity of 
the standard assumptions 1) and 3)
both require an efficient loss of correlations with the initial state.
The characteristic time scale for this loss is given by
the inverse damping rate of the unequal-time correlator 
$\rho^{(eq)}(t,\bp)$ or, similarly, by
$\langle \Phi(t,\bx) \Phi(0,\by) \rangle$. The latter describes
the correlations with the initial time $t_0 = 0$. A vanishing
$\langle \Phi(t,\bx) \Phi(0,\by) \rangle$ corresponds to 
an effective loss of information about details of the initial 
conditions. The characteristic time scale for this loss is well described
by $t_{\rm damp}$ as defined in 
Eq.~(\ref{eq:estimate})~\cite{Berges:2005ai}.\footnote{One may ask 
whether initial correlations
beyond the considered Gaussian initial conditions could change this
situation. However, such initial correlations would be beyond 
a typical transport description based on two-point correlation functions
or effective particle numbers.}

\section{Off-shell transport}

The gradient expansion underlying transport equations is an expansion
in the number of derivatives with respect to
the center coordinates (\ref{eq:center}) and derivatives 
with respect to momenta associated to the relative coordinates
(\ref{eq:relative}).  The LO equations are obtained by 
neglecting ${\cal O}\left(\partial_{X^{\mu}} 
\partial_{p_{\mu}}\right)$
and higher contributions in the gradient expansion, the 
NLO equations by neglecting ${\cal O}\left(\left(\partial_{X^{\mu}} 
\partial_{p_{\mu}}\right)^2 \right)$ etc.\ as described in 
Sec.~\ref{sec:transport}. This expansion is constructed to work
well for on-shell quantities, which are evaluated at the 
peak of the spectral function $\tilde{\varrho} (X^0;\omega,\bp)$
as a function of $\omega$ for given $\bp$.
Away from this peak the contributions coming from derivatives
with respect to momenta are typically not small if the spectral
function is sufficiently narrow. The latter is always the case
for weak enough couplings. Therefore, one may expect 
transport equations to fail for off-shell quantities even
for weak couplings. 

In order to quantify this expectation, we have in mind a
nonequilibrium situation for a given
time $X^0$ larger than $\sim t_{\rm damp}$ but short compared to
characteristic thermal equilibration times. In this case the occupation number 
can still show substantial deviations from a 
thermal distribution though certain properties, such
as a fluctuation-dissipation relation
\beq 
\tilde{F}(\omega,\bp) = \left( n(\omega)+\frac{1}{2}\right)
\tilde{\varrho}(\omega,\bp)\,,
\label{eq:flucdiss}
\eeq
with a non-thermal distribution $n(\omega)$, can become approximately
valid~\cite{FDR}. We assume that the spectral function may be 
approximated by its equilibrium form
\beq
\tilde\varrho(\omega,\bp) = 
\frac{-\tilde\Sigma_{\varrho}(\omega,\bp)}{
\left(-\omega^2+{\bp}^2+M^2+\frac{1}{2}
\tilde\Sigma_+(\omega,\bp)\right)^2
+\left(\frac{1}{2}\tilde\Sigma_{\varrho}(\omega,\bp)\right)^2} \, ,
\label{eq:eqrho}
\eeq
while the non-thermal distribution will be parameterized as 
(cf.~also Eq.~(\ref{eq:init2}))
\beq
n(\omega) = A \exp\left\{-\frac{1}{2\sigma^2}(\omega-\omega_0)^2\right\} \, .
\label{eq:nonthermal}
\eeq
We employ Eq.~(\ref{eq:flucdiss}) with Eqs.~(\ref{eq:eqrho}) and
(\ref{eq:nonthermal}) as a convenient class of approximate parametrizations
of typical nonequilibrium situations at intermediate times as 
have been studied in the previous sections. 
With the self-energies (\ref{eq:SigmaF}) 
and (\ref{eq:Sigmarho}) in Fourier-space  
(cf.~Sec.~\ref{sec:assumptions}) these form a closed set of 
self-consistent equations for $\tilde{F}(\omega,\bp)$ 
and $\tilde\varrho(\omega,\bp)$.
\begin{figure}[t]
\centerline{
\epsfig{file=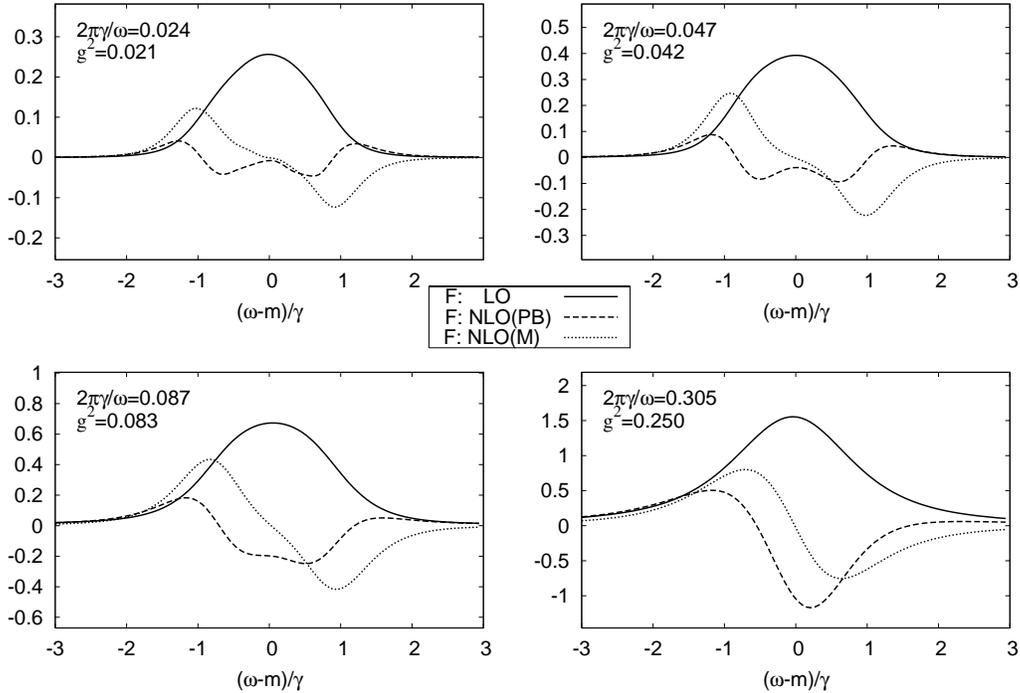,width=14.cm}
\vspace*{-0.6cm}
}
\caption{\small 
The different curves of each graph show the different
contributions 
$\tilde{F} \tilde{\Sigma}_{\varrho} - \tilde{\varrho} \tilde{\Sigma}_F$
(solid line), $\partial_{X^0} M^2 \partial_\omega \tilde{F}$ (dashed)
and $\frac{1}{2}\{ \tilde{\Sigma}_F , \tilde{\varrho}_+\}_{PB} 
+ \frac{1}{2} \{ \tilde{\Sigma}_+ , \tilde{F} \}_{PB}$
(dotted). The first corresponds to the LO contribution in the
gradient expansion, while the latter two add up to the NLO
contribution.
\label{fig:static}
}
\end{figure}

From these solutions we evaluate the LO and NLO expressions 
for $ \partial_{X^0} \tilde{F} \left( X^0;\omega,\bp \right)$
according to Eqs.~(\ref{eq:LOgradF}) and (\ref{eq:NLOF}) in the very same 
way as described in Sec.~\ref{sec:comparison}. The results are shown
in Fig.~\ref{fig:static} for various couplings as a function
of $(\omega - m_R)/\gamma$ for $\bp =0$ with $A =1$ and
$\sigma =m_R$. Here $\gamma$ corresponds to the half width of the
spectral function.
The different curves of each graph show the different
contributions 
$[\tilde{F} \tilde{\Sigma}_{\varrho} 
-\tilde{\varrho} \tilde{\Sigma}_F]/2\omega$
(solid line), $\partial_{X^0} M^2 \partial_\omega \tilde{F}/2\omega$ (dotted)
and $\left[\{ \tilde{\Sigma}_F , \tilde{\varrho}_+\}_{PB} 
+ \{ \tilde{\Sigma}_+ , \tilde{F} \}_{PB}\right]/4\omega$
(dashed) to the transport equation (\ref{eq:NLOF}). 
The first corresponds to the LO contribution in the
gradient expansion, while the latter two add up to the NLO
contribution. One observes that the LO contribution has its maximum
value on shell, i.e.\ for $\omega = m_R$. For small enough couplings
we clearly find that the on-shell NLO corrections are small compared
to LO ones. 

In contrast, Fig.~\ref{fig:static} shows
that for small couplings the NLO corrections have their maximum
contributions off shell. They easily exceed the
LO terms away from $\omega \simeq m_R$ and here a gradient expansion
to low orders cannot be expected to converge well.
When the coupling grows to about $g^2 \simeq 1/4$ the situation changes
qualitatively. Here the Poisson-brackets contribution at NLO becomes
comparable to the LO terms {\em on shell}. Though one expects 
transport equations to become unreliable for sufficiently large
couplings, one observes for our example that NLO starts
to become comparable to LO already for rather weak interactions.
We emphasize that the normalization $g^2 \Phi^4$ we use to define our
coupling reflects rather well the separation between a weakly ($g^2 \ll 1$) 
and strongly coupled regime ($g^2 \gg 1$). To see this we consider
the ratio between
damping rate and oscillation frequency of the unequal-time correlator
$\rho(t,\bp)$, which we denote as $2 \pi \gamma/\omega$. 
Here $2 \pi \gamma/\omega \ll 1$ corresponds to very weakly damped
oscillations, while $2 \pi \gamma/\omega \gg 1$ is overdamped.      
For comparison we display the corresponding values for this
ratio in Fig.~\ref{fig:static}. One observes that they reflect
rather well the coupling strength for the range of couplings considered
here.

\section{Conclusions}

For the considered scalar $g^2 \Phi^4$ theory
we find that the earliest time for the validity of 
transport equations is well characterized by the standard
relaxation or damping time $t_{\rm damp}$ given by Eq.~(\ref{eq:estimate}).
We have done a series of comparisons for various couplings 
and conclude that for times sufficiently large compared to $t_{\rm damp}$
the gradient expansion seems to converge well. There are sizeable NLO 
corrections already for couplings of about $g^2 \simeq 1/4$. 
Nevertheless, one observes that the NLO result can get rather close 
to the ``full'' result for $g^2 \lesssim 1$.
Times shorter than about $t_{\rm damp}$ seem clearly
beyond the range of validity of transport equations
even for weak couplings. This should make them unsuitable
to discuss aspects of apparent early thermalization and stresses
the need to employ proper equations such as (\ref{eq:exactevolF})
and (\ref{eq:exactevolrho}) for initial-value 
problems.

\end{document}